\documentstyle[aps,epsf,multicol]{revtex}

\def\avg#1{\left < {#1} \right >}
\def\etal{{\it et. al. }}
\def\prb{Phys. Rev. B }
\def\prl{Phys. Rev. Lett. }

\begin{document}

\draft
\title{A study of disordered systems with gain: Stochastic Amplification }
\author{Sandeep K. Joshi\cite{jos}}
\address{Institute of Physics, Sachivalaya Marg, Bhubaneswar 751 005, India}

\author{A. M. Jayannavar\cite{amje} }
\address{Institute of Physics, Sachivalaya Marg, Bhubaneswar 751 005, India}

\maketitle

\begin{abstract} 

A study of statistics of transmission and reflection from a random medium
with stochastic amplification as opposed to coherent amplification is
presented.  It is found that the transmission coefficient $t$, for sample
length $L$ less than the critical length $L_c$ grows exponentially with
$L$.  In the limit $L \rightarrow \infty$ transmission decays
exponentially as $\avg{lnt} = -L/\xi$ where $\xi$ is the localization
length. In this limit reflection coefficient $r$ saturates to a fixed
value which shows a monotonic increase as a function of strength of
amplification $\alpha$. The stationary distribution of super-reflection
coefficient agrees well with the analytical results obtained within the
random phase approximation (RPA).  Our model also exhibits the well known
duality between absorption and amplification. We emphasize the major
differences between coherent amplification and stochastic amplification
where-ever appropriate.

\end{abstract}
\begin{multicols}{2}
\narrowtext

With the development of scaling theory of localization our understanding
of the Anderson localization of electrons in disordered systems has
advanced considerably\cite{tvr}. In close analogy with this it was felt
that light waves in a medium with random dielectric constant may also
exhibit localization\cite{sjohn}. In the interaction of light with matter
apart from scattering, absorption plays a very important role. Due to
absorption, light intensity gets strongly attenuated after being scattered
a few times. Presence of gain in the medium not only compensates for this
absorption but also opens up the possibility of observing interesting
phenomenon arising due to interplay of disorder and amplification in
strongly scattering media. This gave birth to the fascinating idea of the
so-called ``random lasers'' which are essentially mirror-less
lasers\cite{Nkupp}. The obvious importance of this in the field of laser
instrumentation together with viability of multiple scattering experiments
using powdered laser crystals or micro-particle suspensions in laser dyes
fueled an intense research activity in the field in recent
years\cite{expts}. The different techniques employed in the theoretical
investigations include the time-independent studies of reflection and
transmission using invariant imbedding\cite{Nkupp,jhein,pusti,rrbd}
equations and transfer matrix technique\cite{asen,zhang,abhi,joshi,jiang}
for one-dimensional systems, Fokker-Planck equation for many channel case
\cite{beenak,pass,misir} and time dependent diffusion equation
approach\cite{zyuzin}. In these studies the gain or
amplification(absorption) is modeled by introducing an imaginary part into
the dielectric constant in case of classical (electro-magnetic) waves in
random medium or into the site energy for the tight-binding
models\cite{abhi,joshi}. This amounts to making the potential complex
leading to a non-Hermitian Hamiltonian\cite{Nkupp,jhein,pusti,asen,
zhang,abhi,joshi,jiang,beenak,pass,misir,zyuzin,rubio,amj,qli}. The
resulting amplification is referred to as coherent amplification.

The presence of imaginary potential gives rise to many a counter-intuitive
features and some interesting results. Due to the real to complex mismatch
experienced by the incident waves in the vicinity of the amplifying medium
there is an enhanced back reflection. Thus the medium acts as an amplifier
as well as a reflector. This dual role implies that amplification without
reflection is not possible\cite{rubio,amj}. In fact in the limit of the
strength of imaginary part of the potential becoming infinite, reflection
probability tends to unity, i.e., the amplifier acts as a perfect
reflector\cite{abhi,joshi}. The stationary distribution from 
such a disordered coherently amplifying medium, calculated under the
assumption of random phase approximation using invariant imbedding method
is given by\cite{Nkupp}


\begin{eqnarray}
P_s(r) & = & \frac{|D|exp(-\frac{|D|}{r-1})}{(r-1)^2}~~~for~~ r \geq 1
\label{psr} \\
       & = &  0~~~~~~~~~~~~~~~~~                     ~~~for~~ r < 1
\nonumber
\end{eqnarray}

\noindent Here $D$ is proportional to $V_i/W$, $V_i$ and $W$ being the
strength of imaginary part of the potential and disorder respectively. The
distribution peaks at a single value of $r>1$ which keeps shifting to
higher values of $r$ with increasing amplification. On the contrary, the
exact distribution obtained numerically\cite{joshi} exhibits a double
peaked structure for strong amplification or disorder and in the limit of
perfect reflector ($V_i\rightarrow\infty$) becomes a delta function at
$r=1$ in consistency with the physics of the problem. The transmission
coefficient is also known to exhibit equally surprising features. Despite
amplification, in the limit of sample length $L\rightarrow\infty$ the
transmission falls off exponentially as a function of $L$ with a well
defined localization length $\xi$ not only when there is disorder along
with but also in case of a perfectly periodic amplifying
medium\cite{asen,zhang,abhi,joshi,jiang,beenak,pass,misir}. Moreover, the
rate of exponential decay is the same as that of an absorbing medium,
i.e., $\xi(V_i)=\xi(-V_i)$. This symmetry, which has been referred to as
duality between absorption and amplification in the literature follows
from the time-independent wave equation\cite{pass}. Recently, Soukoulis
et. al. have claimed that this paradoxical result is a mere artifact of
time-independent equations. Based on the analysis of the geometric series
for the transmission amplitude they argue that depending on the gain
parameter the series becomes divergent beyond a particular length
indicating the absence of stable time-independent solutions\cite{qli}.
However, it is to be noted that the exact calculation of the transmission
coefficient (as a scattering problem) as a function of length for an
ordered lasing medium does not show any divergence. As a function of
length transmission coefficient exhibits a maximum both for ordered as
well as disordered lasing systems.

The interesting features discussed above seem to be a fallout of the fact
that the phase coherence of the waves in a medium thus modeled is
maintained in spite of the presence of amplification justifying the term
coherent amplification. In the present work we endeavor to study a
different model for amplification in the framework of time-independent
wave equations which gives qualitatively different results for statistics
of super-reflection in the strong amplification regime. The model was used
to study stochastic absorption also\cite{sahoo}. In this simple minded
model a chain of uniformly spaced random strength delta function
scatterers act as the disordered medium. A negative attenuation constant
per unit length $\alpha$ characterizes the stochastic amplification. The
free propagator acquires a factor $exp(\alpha a)$ for each trip while
traversing the free region of length $a$ in between the
scatterers\cite{datta}. We find that this method of incorporating
amplification avoids the additional reflections and resonances observed in
case of coherent amplification as mentioned above. The continuum limit of
our model yields the same Langevin equation for reflection amplitude
$R(L)$ as obtained by Pradhan\cite{pp,jo_amj_unpub} for the case of
stochastic absorption except that $\alpha$ is to be replaced by
$-2\alpha$. The stationary solution of the Fokker-Planck equation for the
probability distribution of reflection coefficient $P_s(r)$ in case of
lasing medium obtained within RPA, however, is the same as given by Eq.
\ref{psr} with the $D$ in the expression now proportional to $\alpha/W$.
The numerically obtained exact stationary distribution of reflection
coefficient agrees well with the analytical result for small values of
$\alpha$ but in larger parameter space as compared to the coherent
amplification case\cite{joshi,sahoo}. With increasing $\alpha$ the
stationary value $\avg{lnr}_s$ keeps increasing monotonically as against
the non-monotonic behavior observed in case of coherent amplification.  
For coherent amplification $\avg{lnr}_s$ approaches zero in the $\alpha
\rightarrow \infty$ limit as the amplifier then acts as a perfect
reflector. One does not expect this from a realistic amplifying medium.
Our phenomenological model rectifies this problem. Although the properties
of reflection coefficient for stochastic amplification are qualitatively
contrasting, the properties of transmission show some similarities with
those observed in case of coherent amplification. In the present model the
transmission decays exponentially with the sample length $L$ with a well
defined localization length $\xi$. This localization length turns out to
be equal to the one obtained for stochastic absorption, i.e., we find that
this model also exhibits the dual symmetry.  However, there are some minor
differences in regard to the critical system length $L_c$ beyond which the
transmission starts decaying. In the following we begin by briefly
describing the model and numerical procedure. We then present results and
discussions and finally we give some conclusions.

We take recourse to the transfer matrix technique for our numerical
calculations\cite{abrahams,cbasu}. As mentioned earlier the chain of
random strength delta function scatterers spaced at regular intervals of
length $a$ plays the role of a disordered medium. The transfer matrix for
$j^{th}$ delta function having strength $q_j$ is

$${\sf M_j}~=~
\left (
\begin{array}{cc}
1-iq_j/2k & -iq_j/2k \\
iq_j/2k & 1+iq_j/2k \\
\end{array} 
\right ) $$

\noindent The $q_j$'s are chosen from a flat distribution between $-W/2$
and $W/2$, i.e., $P(q_j)=1/W$ where $W$ is the strength of the disorder.
We choose to work with units such that $\hbar$ and $2m$ are unity and
therefore $E=k^2$ is the energy of the incident wave. $W$ and $\alpha$ are
scaled to make them dimensionless. Stochastic amplification is introduced
in the free propagation medium through the transfer matrix\cite{sahoo}

$$ {\sf X}~=~
\left ( \begin{array}{cc}
e^{ik+\alpha} & 0 \\
0 & e^{-ik-\alpha} \\
\end{array} \right ). $$

\begin{figure}
\protect\centerline{\epsfxsize=2.5in \epsfbox{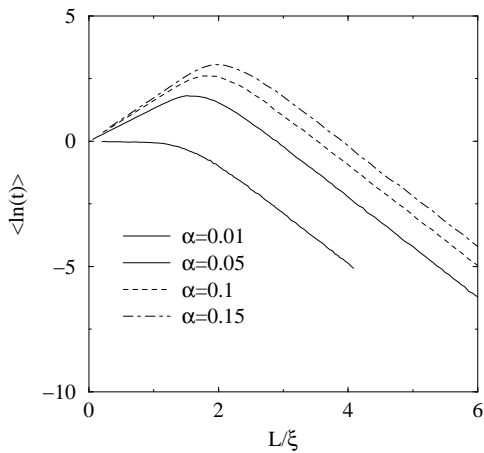}}
\caption{Average of logarithm of transmission coefficient $t$ versus
$L/\xi$ for different values of amplification strength $\alpha$ as
indicated in the figure and disorder strength $W=1.0$.}
\label{lntvsl} 
\end{figure}

\begin{figure}
\protect\centerline{\epsfxsize=2.5in \epsfbox{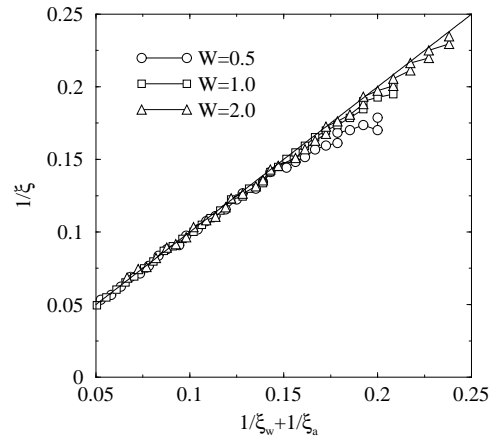}}
\caption{$1/\xi$ versus $1/\xi_w+1/\xi_a$, where $\xi_w$ is the
localization length for passive (non-amplifying) disordered system,
$\xi_a$ is the gain length for ordered ($W=0$) amplifying system and $\xi$
is the localization length for amplifying disordered system ($W=1.0$. }
\label{xiw}
\end{figure}

\begin{figure}  
\protect\centerline{\epsfxsize=2.5in \epsfbox{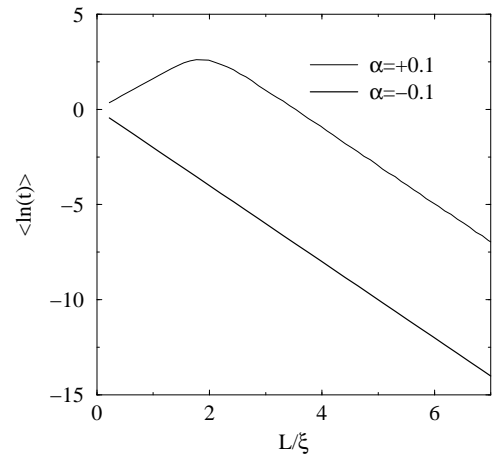}}
\caption{Variaion of average of logarithm of transmission coefficient with 
sample length $L$ for $\alpha = 0.1$ (amplification) and $\alpha = -0.1$ 
(absorption). For both the cases the disorder strength $W=1.0$.}
\label{duality}  
\end{figure}

The total transfer matrix for the stochastically amplifying medium is then
obtained by recursive multiplication of these matrices i.e., $$M~=~{\sf
M_LX....XM_2XM_1}.$$ One can then calculate the reflection and
transmission coefficient from the matrix $M$ by using the well-know
formulae\cite{cbasu} $$ R~=~-\frac{M(2,1)}{M(2,2)} $$ and
$$T~=~-\frac{{\sf det}M}{M(2,2)}.$$ Needless to say that since we are
working with non-Hermitian systems $r+t\neq1$. In the results presented
here we have considered at least 10,000 realizations for calculating the
various distributions and averages and the incident energy is chosen to be
$E=k^2=1$ unless specified otherwise. We consider chains of length about
10 times the localization length in order to evaluate the stationary
distributions. The distributions did not evolve on increasing length
beyond $10\xi$. To show the exponential scaling of transmission
coefficient with length $L$ of the sample we plot $\avg{lnt}$ versus
$L/\xi$ in Fig. \ref{lntvsl} for $W=$1.0 and different values of
amplification strength $\alpha$ ($\alpha=$0.01,0.05,0.1,0.15). From the
figure it is evident that there is a critical length $L_c$ of the system
upto which the transmission grows and beyond that the exponential decay
takes over. For $L<L_c$ the transmission increases with sample length as
$\avg{lnt}=(1/\xi_a-1/\xi_w)L$ as found in earlier
studies\cite{jhein,jiang}. Here $\xi_a \sim 1/\alpha$ is the gain length
characterizing the exponential growth of the transmission in absence of
disorder and presence of amplification only. $\xi_w=96k^2/W^2$ is the
localization length in the presence of disorder alone\cite{econ}. In the
limit $W\rightarrow 0$, $\avg{lnt}\sim L/\xi_a$ and $L_c \rightarrow
\infty$ i.e., in absence of disorder transmission keeps growing with
sample length. In contrast to this, for coherent amplification $L_c
\rightarrow L^0_c$, a fixed value for $W \rightarrow 0$. This is due to
the fact that even in the ordered case coherent amplification leads to
backscattering, i.e., reflection. In the present case of stochastic
amplification for an ordered system transmission coefficient grows
exponentially with length. The presence of disorder induces coherent
backscattering leading to the exponential decay (localization) of the
transmission in the asymptotic limit in spite of amplification. For
lengths greater than $L_c$, i.e., $L\gg L_c$, the behavior of $\avg{lnt}$
versus $L$ is given by $\avg{lnt}=-L/\xi$, where $\xi$ is the localization
length scaled with respect to inter-scatterer spacing $a$. The plot of
$1/\xi$, calculated numerically by changing $\alpha$ for various values of
disorder strength, versus $(1/\xi_a + 1/\xi_w)$ shown in Fig.  \ref{xiw}
suggests the scaling relation $1/\xi=(1/\xi_a + 1/\xi_w)$ for the
localization length in presence of both disorder and amplification. The
value of $\xi$ thus obtained was used for the plot of Fig. \ref{lntvsl}.
We note that this relation is identical to one obtained for the
localization length in case of a stochastically absorbing medium, i.e.,
$\xi(\alpha,W) = \xi(-\alpha,W)$\cite{sahoo}. This gives the numerical
evidence for the fact that our model also exhibits the much debated
duality between absorption and amplification. Although the same duality is
observed in case of coherent amplification and absorption it is important
to realize that in the present case the duality relation holds only for
non-zero disorder strength ($W\neq0$) whereas in the former it was valid
even for $W=0$. To explicate the point of dual symmetry we have shown the
plot of $\avg{lnt}$ versus $L$ for $\alpha=\pm 0.1$ in Fig. \ref{duality}.
From the figure we see that the asymptotic slope of the transmission
coefficient for absorption as well as amplification ( $L \gg L_c$) are
equal.

\begin{figure}
\protect\centerline{\epsfxsize=2.5in \epsfbox{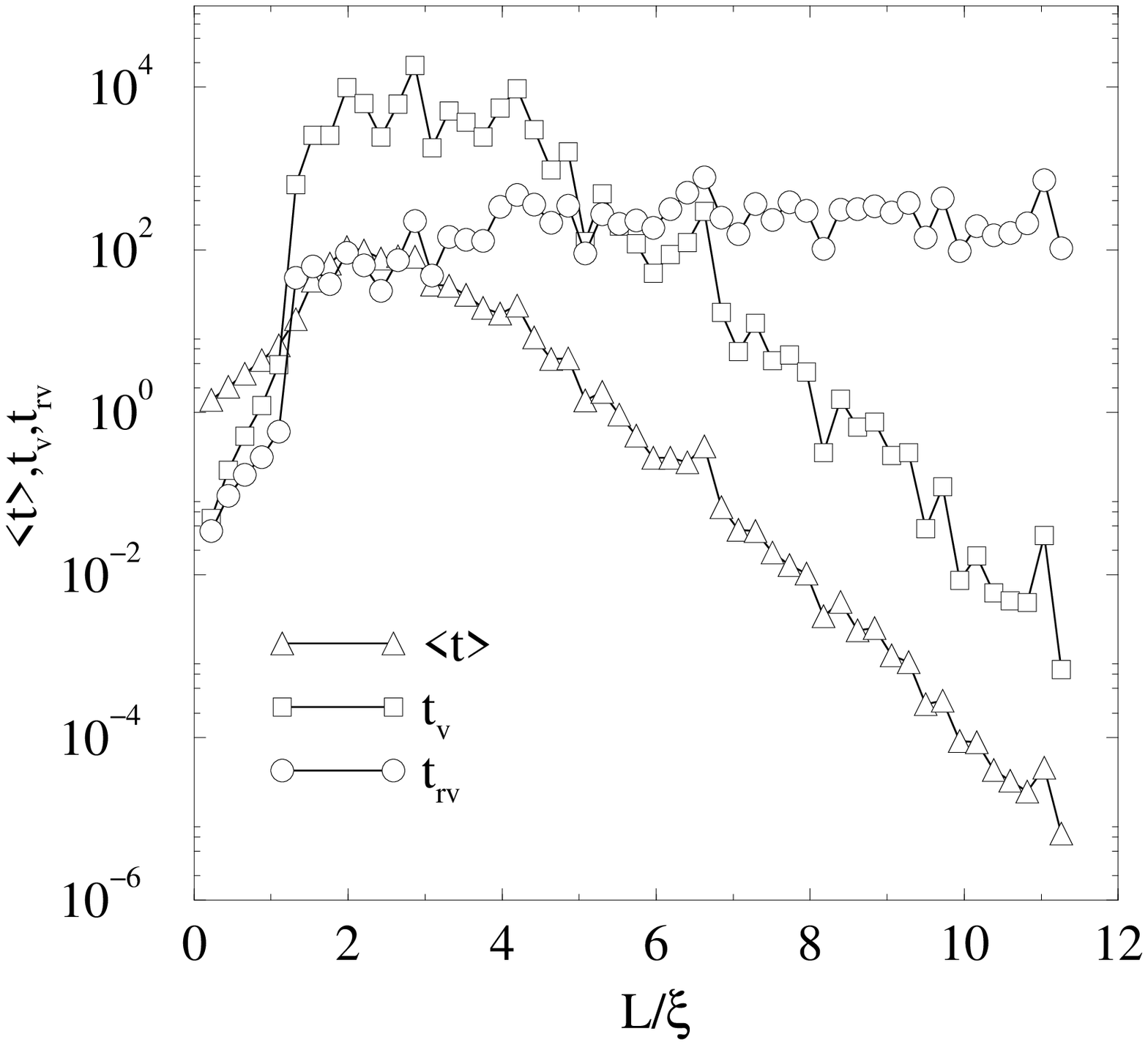}}
\caption{Average transmission coefficient $\avg{t}$, rms variance of $t$ and
rms relative variance of $t$ as a function of $L/\xi$ for fixed disorder
$W=1.0$ and amplification $\alpha=0.01$.}
\label{fluct}
\end{figure}

\begin{figure}
\protect\centerline{\epsfxsize=2.5in \epsfbox{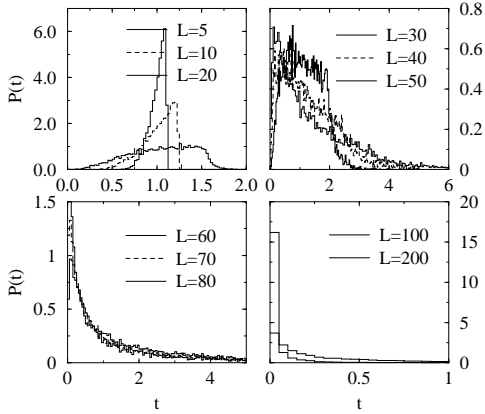}}
\caption{Distribution of transmission coefficient $t$ from a disordered
amplifying system with $W=1.0$ and $\alpha=0.01$ at different lengths
$L$. Localization length for these parameters is $\xi \sim 50$.}
\label{tdist}
\end{figure}

\begin{figure} 
\protect\centerline{\epsfxsize=2.5in \epsfbox{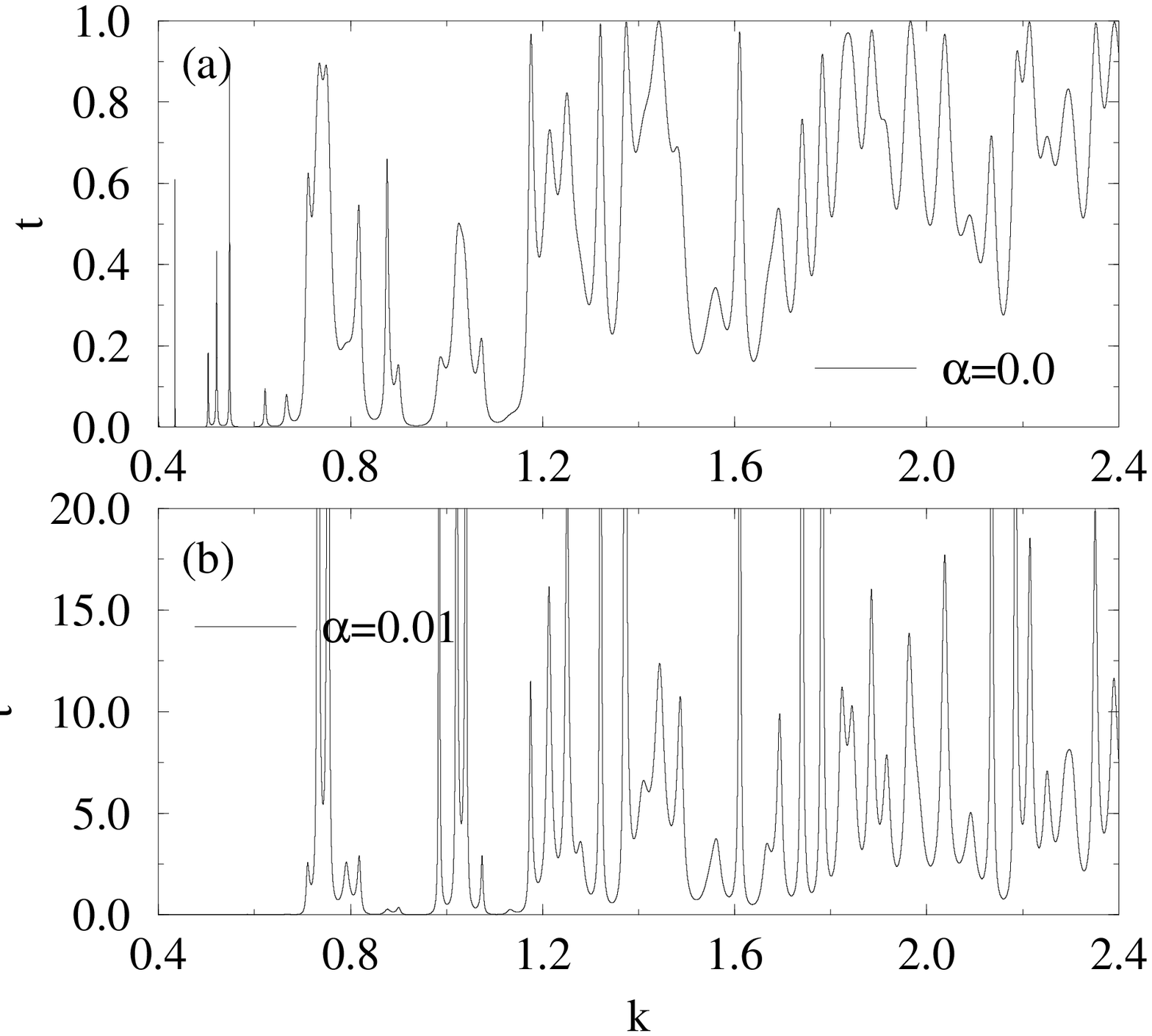}}
\protect\centerline{\epsfxsize=2.5in \epsfbox{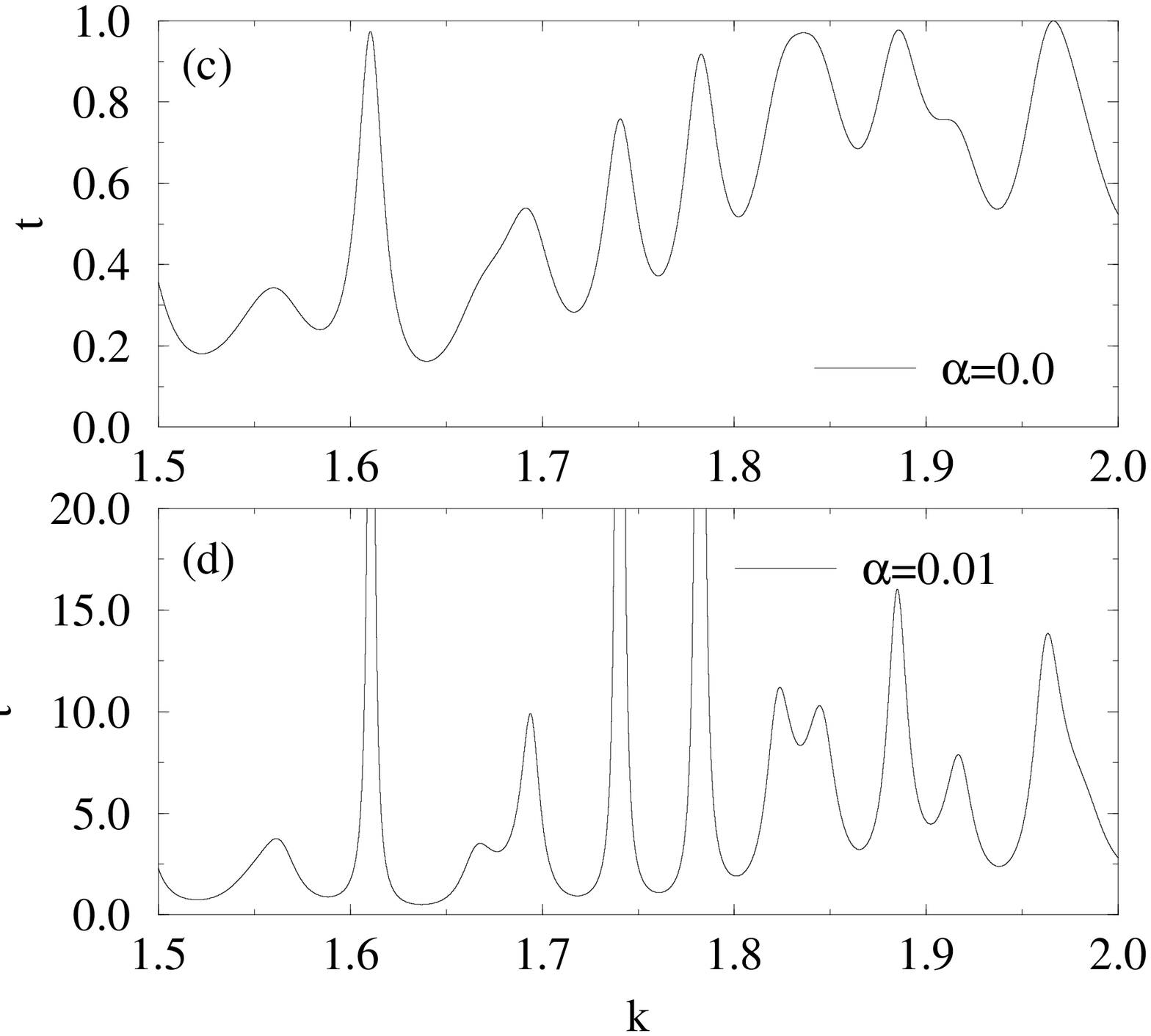}}
\caption{Transmission coefficient as a function of incident wavenumber $k$
for (a) disordered non-amplifying sample ($W=1.0,\alpha=0.0$) of length $L=100$ and
(b) disordered amplifying sample ($W=1.0,\alpha=0.01$) of length
$L=100$. (c) and (d) show a magnified view of the region between $k=1.5$
to $k=2.0$ of (a) and (b) respectively. } 
\label{tvsk} 
\end{figure}

The log-scale plot shown in Fig. \ref{fluct} illustrates the dependence of
the avarage transmission $\avg{t}$, the root mean squared (rms) variance
of $t$, namely $t_v=\sqrt{\avg{t^2}-\avg{t}^2}$ and the rms relative
variance $t_{rv}=t_v/\avg{t}$ on the sample length $L$. For $L<\xi$, rms
variance of $t$ is less than the average value of $t$, i.e., $t_v<\avg{t}$
and consequently rms relative variance is less than unity ($t_{rv}<1$).
However, for $L>\xi$ the rms variance of $t$ exceeds the average
transmission ($t_v>\avg{t}$) and rms relative variance crosses unity
($t_{rv}>1$) indicating the onset of non-self-averaging (NSA) nature of
transmission\cite{fuku,nkuamj}. Therefore for samples of length $L>\xi$
the average transmission over an ensemble of macroscopically identical
samples is dominated over by the sample-to-sample fluctuations in $t$. The
sensitivity of the transmission coefficient to the exact spatial
realization of the impurities gets highly enhanced. A look at the plot of
the distribution of the transmission coefficient at different lengths
shown in Fig. \ref{tdist} helps to shed more light on this. The parameters
for the figure are $\alpha=$0.01 and disorder strength $W$=1.0.
Corresponding localization length is $\xi \sim$50.0. For sample lengths
(see $L=$5) much smaller than the localization length ($L<<\xi$), though
the resonant transmission dominates, the distribution peaks at a value of
$t$ close to unity because amplification strength is small. On increasing
the length $L$ further ($L=$10,20) the peak shifts to higher values of
$t$. When the sample length becomes comparable to the localization length
($L\sim\xi\sim 50$), multiple reflections start dominating. Due to these
multiple reflections time spent inside the medium increases thereby
enhancing the amplification. This broadens the distribution and tails
start appearing. This, however, does not continue indefinitely with
increasing length. As seen earlier finally in the long length limit ($L\gg
\xi$) transmission decays to zero exponentially with length. This reflects
in the distributions of $t$ also. The peak of the distribution starts
shifting to values of $t$ much less than one and simultaneously develops
long tails. These tails are a clear signature of the sample-to-sample
fluctuations which tender the non-self-averaging nature to the
transmission coefficient. The origin of these tails can presumably be
traced back to the so-called Azbel resonances observed in case of passive
disordered media\cite{azbel}. Since we have amplification in addition to
disorder it would be naive to expect no change whatsoever in the structure
of these resonances. Thus a study of resonances in our model is in order.
For the purpose we assume that the ergodic hypothesis of relation between
the ensemble fluctuations and the fluctuations for a given sample as a
function of the chemical potential holds true in the present case. Plot of
Fig. \ref{tvsk} shows the variation of transmission coefficient with the
wave vector $k$ for a single realization of the disorder, both in the
absence and in the presence of amplification. The plot reveals that
presence of amplification modifies the resonances observed in case of
passive medium in two ways namely, the value of transmission at the
resonances is either enhanced ($t>1$) or reduced ($t<1$) depending upon
the sample length $L$ and the widths of the resonance peaks is reduced in
case of the amplifying medium. Unresolved or closely spaced resonance
peaks in the case of passive disorder are resolved due to the narrowing
effect of amplification. This also can be interpreted in the following
manner: two nearby overlapping resonances may have charateristically
different delay times\cite{joshid} or dwell times\cite{butt}. Due to this
enhancement of the transmittance by the amplification will be very
different and hence the resolution. This clearly indicates that
amplification does not give rise to any new resonances and only modifies
the existing resonances vis-a-vis their strength and resolution. We have
verified the above observation for many different macroscopic realizations
of the underlying disorder. This is in contrast to the case of coherent
amplification where new resonances can appear due to the reflecting nature
of imaginary potentials. Such complex potentials can give additional phase
shifts leading to new resonances in addition to those seen in case of the
same disordered realization but in the absence of imaginary potential.
 
\begin{figure}
\protect\centerline{\epsfxsize=2.5in \epsfbox{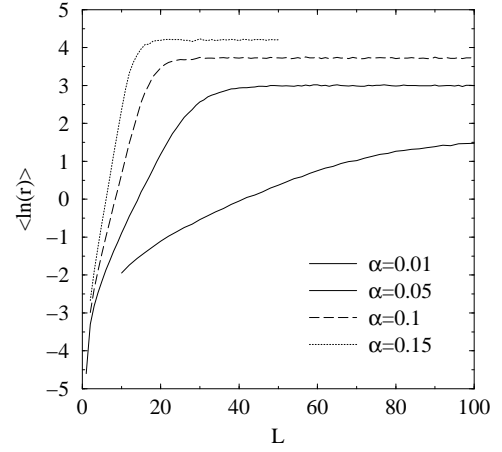}}
\caption{Average of logarithm of reflection coefficient versus sample
length for a fixed value of disorder $W=1.0$ and different values of 
amplification $\alpha$ indicated in the figure. }
\label{lnrvsl}
\end{figure}

We shall now discuss about the statistics of reflection coefficient which
exhibits properties radically different from those observed for coherent
amplification in the limit of strong amplification. Fig. \ref{lnrvsl}
shows the variation of $\avg{lnr}$ with $L$ for a fixed value of disorder
strength $W=$1.0 and various values of amplification strength $\alpha$ as
indicated in the figure. As a function of length $\avg{lnr}$ increases
monotonically.  In the long length limit $L\gg \xi$, $\avg{lnr}$ saturates
to a certain positive value $\avg{lnr}_s$ depending upon the value of
disorder strength $W$ and amplification strength $\alpha$. At any length
we notice that reflection (super-reflection) increases with increasing
$\alpha$.

\begin{figure}
\protect\centerline{\epsfxsize=2.5in \epsfbox{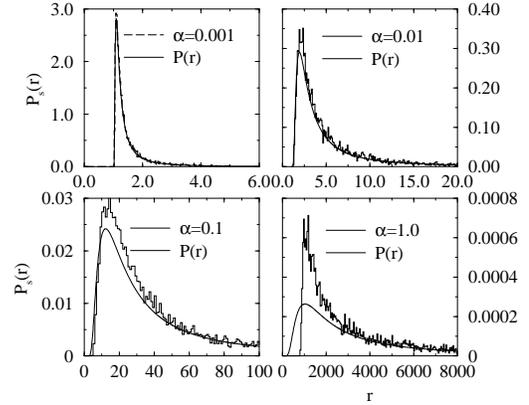}}
\caption{Stationary distribution of reflection coefficient for a fixed
value of disorder strength $W=1.0$ and different values of amplification
strength $\alpha$. The thick line shows the single parameter fit of
analytical expression Eqn.\ref{psr} with (a) $D=0.197$ for $\alpha=0.001$,(b) 
$D=1.92$ for $\alpha=0.01$ and (c) $D=20.53$ for $\alpha=0.1$.}
\label{rdist}
\end{figure}

\begin{figure}
\protect\centerline{\epsfxsize=2.5in \epsfbox{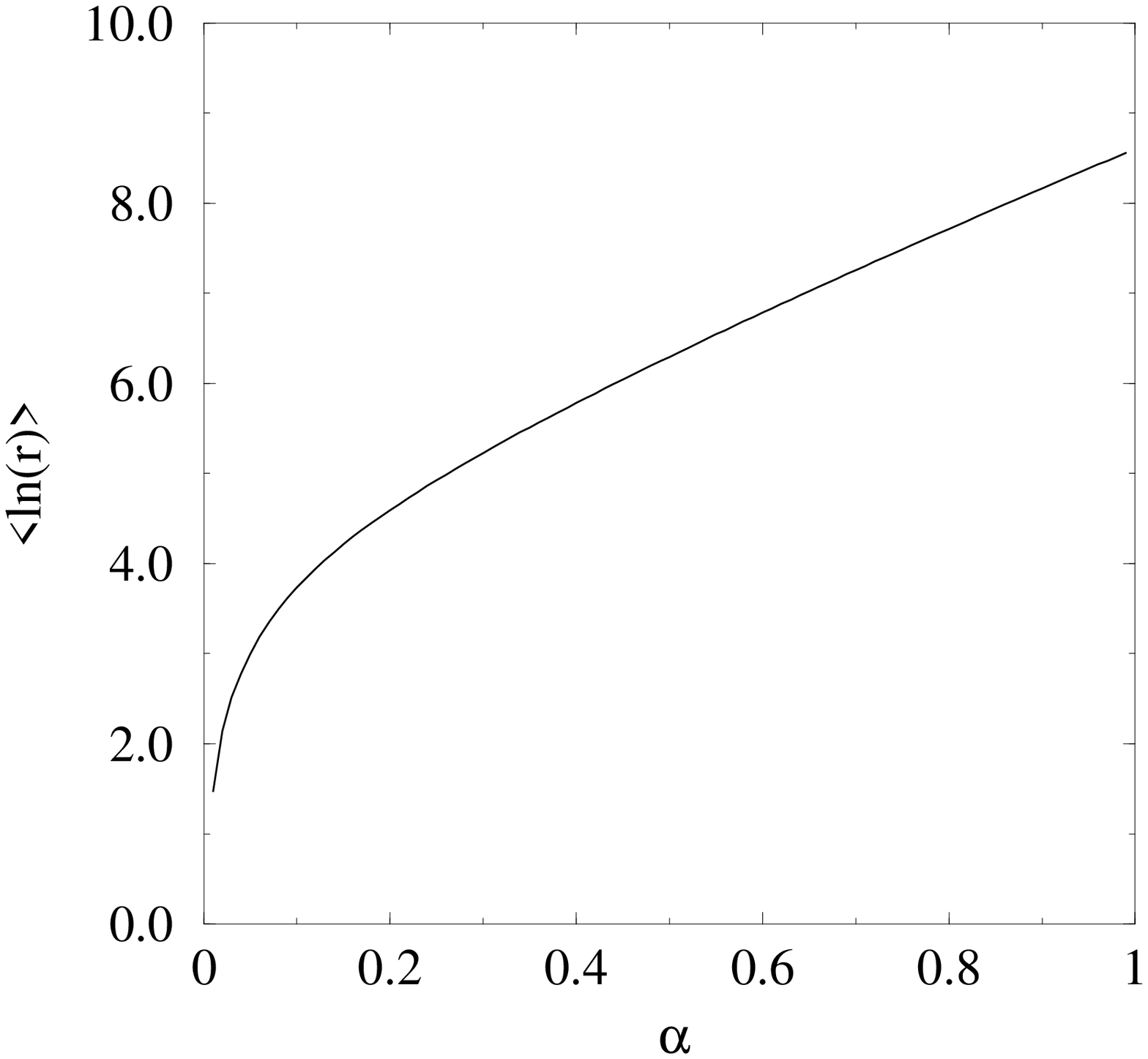}}
\caption{Average value of logarithm of reflection coefficient
($\avg{lnr}$) versus amplification strength $\alpha$ for $W=1.0$ and
$L/\xi=10$.}
\label{lnrvsalpha}
\end{figure}

In Fig. \ref{rdist} we show the stationary distribution of reflection
coefficient $P_s(r)$ for different values of amplification strength
$\alpha$. For small values of $\alpha$ ($\alpha=0.001$) the distribution
peaks at small values of $r$ close to unity. On increasing $\alpha$ the
peak shifts to larger values of $r$. The single parameter fit obtained
using the analytical expression given in Eq. \ref{psr} is shown by a thick
line. On increasing $\alpha$ further ($\alpha>0.1$) the peak of the
distribution continues to shift to higher values of $r$. Although the fit
is no more valid, the qualitative features of the distribution, i.e.,
single prominent peak and long tail, do not change. We have observed in
our numerical calculations that in this parameter range the distribution
of phase of reflected wave shows two distinct peaks indicating breakdown
of RPA. This is in striking contrast to the perfect reflector behavior
observed for the case of coherent amplification\cite{joshi} where on
increasing amplification beyond a certain strength the distribution starts
moving to lower values of $r$ and in the limit $V_i\rightarrow \infty$,
$P_s(r)\rightarrow \delta(r-1)$. In the intermediate regime of $V_i$, the
$P_s(r)$ in the coherent amplification exhibits a double peak structure
reminiscent of the additional reflections introduced by the imaginary
potential (see Ref. \onlinecite{joshi} for details). We do not have any
such additional reflections arising due to amplification in the case of
stochastic amplification. This is also reflected from the plot of
$\avg{lnr}_s$ versus amplification strength $\alpha$ shown in Fig.
\ref{lnrvsalpha}. Recall that the $\avg{lnr}_s$ was a non-monotonic
function of $\alpha$ in case of coherent amplification. In contrast here
we find that $\avg{lnr}_s$ is a monotonically increasing function of
$\alpha$ in accordance with the physical expectations from this model.

In summary, we have explored the effect of disorder on the stochastic
amplification by numerically studying the statistics of transmission and
reflection. It is found that localization occurs solely due to the
presence of disorder. The critical length $L_c$ goes to infinity as
disorder strength is reduced to zero. The transmittance for absorbing as
well as amplifying medium falls off exponentially with length in the
asymptotic limit with same localization length depending on the values of
$|\alpha|$ and $W$. Thus the model exhibits duality between absorption and
amplification. The transmittance is found to be NSA for $L > \xi$.
Stochastic amplification does not introduce any new resonances in the
behavior of transmittance as a function of energy. The average value of
logarithm of reflection coefficient shows monotonic increase with
increasing amplification strength $\alpha$ in contrast to the behavior
observed in case of coherent amplification. The numerically calculated
stationary distribution agrees well with the analytical result in a larger
parameter space. For strong amplification regime the distribution still
retains the same qualitative features with peak shifting to higher values
of reflection. Thus there is no additional reflection due to presence of
amplification as against the observation in case of coherent
amplification.

\end{multicols}

\begin{references}

\bibitem[\dagger]{jos}e-mail: joshi@iopb.res.in

\bibitem[\star]{amje}e-mail: jayan@iopb.res.in

\bibitem{tvr}P. A. Lee and T. V. Ramakrishnan, Rev. Mod. Phys. {\bf 57}, 287 
(1985).

\bibitem{sjohn} See also S. John, in {\it Scattering and Localization of
Classical Waves in Random Media}, edited by P. Sheng World
Scientific,Singapore, 1990; also Phys. Today 44 5, 32 1991.

\bibitem{expts} N. M. Lawandy \etal, Nature {\bf 368}, 436 (1994);
D. S. Wiersma, M. P. van Albada and Ad Lagendijk, \prl {\bf 75}, 1739 (1995);
M. Siddique et. al., Opt. Lett. {\bf 21}, 451 (1996); 
B. Raghavendra Prasad et. al., Appl. Opt. {\bf 36}, 7718 (1997).

\bibitem{Nkupp} P. Pradhan and N. Kumar, \prb {\bf 50}, 9644 (1994).

\bibitem{jhein} J. Heinrich , \prb {\bf 52}, 4156 (1995).

\bibitem{pusti} V. Freilikher, M. Pustilnik, and I. Yurkevich, \prb {\bf
56}, 5974 (1997); V. Freilikher, M. Pustilnik and I. Yurkevich, \prl
{\bf 73}, 810 (1994).

\bibitem{rrbd} R. Rammal and B. Doucot, J. Phys. (Paris) {\bf 48}, 509
(1987).

\bibitem{asen} A. K. Sen, Mod. Phys. Lett. B {\bf 10}, 125 (1996).

\bibitem{zhang} Z. Q. Zhang, \prb {\bf 52}, 7960 (1995).

\bibitem{abhi} A. K. Gupta and A. M. Jayannavar, \prb {\bf 52}, 4156 (1995).

\bibitem{joshi} S. K. Joshi and A. M. Jayannavar, \prb {\bf 56}, 12038
  (1997); S. K. Joshi and A. M. Jayannavar, Int. Jou. Mod. Phys. {\bf
    B12}, 1555 (1998).  

\bibitem{jiang} X. Jiang and C. M. Soukoulis, \prb {\bf 59}, 6159 (1999).

\bibitem{beenak} C. W. J. Beenakker, J. C. J. Paasschens and P. W. Brouwer, 
\prl {\bf 76}, 1368 (1996).

\bibitem{pass} J. C. J. Paasschens, T. Sh. Misirpashaev and C. W. J. Beenakker,
\prb {\bf 54}, 11887 (1996).

\bibitem{misir} T. Sh. Misirpashaev, J. C. J. Paasschens and C. W. J. 
Beenakker, Physica A {\bf 236}, 189 (1997).

\bibitem{zyuzin} A. Yu. Zyuzin, \pre {\bf 51}, 5274 (1995).

\bibitem{rubio} A. Rubio and N. Kumar, \prb {\bf 47}, 2420 (1993).

\bibitem{amj} A. M. Jayannavar, \prb {\bf 49}, 14718 (1994).

\bibitem{qli} X. Jiang, Q. Li and C. M. Soukoulis, \prb {\bf 59}, R9007 (1999).

\bibitem{sahoo} S. K. Joshi, D. Sahoo and A. M. Jayannavar, submitted.

\bibitem{datta}S. Datta, {\it Electron transport in mesoscopic
systems}, Cambridge University Press, 1995, pg. 260.

\bibitem{pp} P. Pradhan, preprint cond-mat/9807312; P. Pradhan, Thesis,
Indian Institute of Science, Bangalore, India, (1997), unpublished.

\bibitem{jo_amj_unpub} S. K. Joshi and A. M. Jayannavar, unpublished.

\bibitem{abrahams} E. Abhrams and M. J. Stephen, J. Phys. C: Solid
  St. {\bf 13}, L377 (1980); B. S. Andereck and E. Abrahams,  {\it
    ibid.} L383 (1980). 

\bibitem{cbasu} Y. Liu and K. A. Chao, \prb {\bf 34}, 5247 (1986);
  P. K. Thakur, C. Basu, A. Mookerji and A. K. Sen,
  J. Phys. Condens. Matter {\bf 4}, 6095 (1992).

\bibitem{econ} E. N. Economou, {\it Green's Functions in Quantum
    Physics}, 2nd ed. (Springer-Verlag, Berlin, 1983), p. 174. 

\bibitem{azbel} M. Ya. Azbel, Solid State Commun. {\bf 45}, 527 (1983).



\bibitem{fuku} P. A. Lee, A. D. Stone and H. Fukuyama, \prb {\bf 35}, 1039
(1987).

\bibitem{nkuamj} N. Kumar and A. M. Jayannavar, \prb {\bf 32}, 3345
(1985)



\bibitem{joshid} S. K. Joshi, Abhijit Kar Gupta and A. M. Jayannavar, \prb
{\bf 58}, 1092 (1998); S. K. Joshi and A. M. Jayannavar, Solid State
Commun. {\bf 106}, 363 (1998) ; {\it ibid} {\bf 111}, 547 (1999) ; A. M.
Jayannavar, G. V. Vijayagovindan, N. Kumar, Z. Phys. B {\bf 75}, 77
(1989).

\bibitem{butt} M. B\"uttiker, \prb {\bf 27}, 6178 (1983).


\end{references}
\end{document}